\documentclass{ws-procs975x65}

\begin{document}

\title{SEIBERG-WITTEN INSTABILITY OF VARIOUS TOPOLOGICAL BLACK HOLES}

\author{YEN CHIN ONG$^{\text{1,3}*}$ and PISIN CHEN$^{\text{1-4}}$}

\address{$^1$Graduate Institute of Astrophysics, National Taiwan University, Taipei 10617, Taiwan.\\
 $^2$Department of Physics, National Taiwan University, Taipei 10617, Taiwan.\\
$^3$Leung Center for Cosmology and Particle Astrophysics, \\National Taiwan University, Taipei 10617, Taiwan.\\
$^4$Kavli Institute for Particle Astrophysics and Cosmology, \\SLAC National Accelerator Laboratory, Stanford University, Stanford, CA 94305, U.S.A.\\
$^*$E-mail: ongyenchin@member.ams.org}

\begin{abstract}
We review the Seiberg-Witten instability of topological black holes in Anti-de Sitter space due to nucleation of brane-anti-brane pairs. We start with black holes in general relativity, and then proceed to discuss the peculiar property of topological black holes in Ho\v{r}ava-Lifshitz gravity -- they have instabilities that occur at only finite range of distance away from the horizon. This behavior is not unique to black holes in Ho\v{r}ava-Lifshitz theory, as it is also found in the relatively simple systems of charged black hole with dilaton hair that arise in low energy limit of string theory.
\end{abstract}


\bodymatter

\section{Introduction: Seiberg-Witten Instability}
In view of the various applications in the Anti-de-Sitter/Conformal Field Theory (AdS/CFT) correspondence, black hole solutions in AdS have received much attention in the literature\cite{Mann, Birmingham}. In the original formulation by Maldacena\cite{Maldacena}, it is conjectured that type IIB superstring theory in $\text{AdS}_5 \times S^5$ is dual to $\mathcal{N}=4$ U($N$) super-Yang-Mills theory in $(3+1)$-dimensions. This has been generalized to other dimensions. In principle, the geometry of spacetimes can be affected by the presence of branes in asymptotically locally AdS spacetime. Seiberg and Witten showed quite generically that if a certain function (the brane action) defined on the Wick-rotated spacetime becomes negative, the spacetime becomes unstable\cite{SW, Kleban}. 

Specifically, given a BPS (Bogomol'ny-Prasad-Sommerfield) brane $\Sigma$ in Wick-rotated $d$-dimensional spacetime, the Seiberg-Witten brane action in the appropriate unit is a function of radial coordinate $r$ defined by
$
\mathcal{S}[\Sigma(r)] = \mathcal{A}(\Sigma) - (d-1)\mathcal{V}(\Sigma),
$
where $\mathcal{A}$ is the area of $\Sigma$ and $\mathcal{V}$ its volume.
The instability is caused by nucleations of brane-anti-brane pairs at the regions where the brane action is negative, a phenomenon analogous to Schwinger pair-production in strong electromagnetic fields\cite{Barbon}. The presence of copious amount of branes alters the geometry. Prior to Seiberg and Witten's work, Maldacena, Michelson, and Strominger already pointed out that various AdS geometries are prone to such drastic changes, which they referred to as ``fragmentation''\cite{MMS}.

It is worth emphasizing that Seiberg-Witten instability applies to any $d$-dimensional spacetime ($d \geq 4$) with an asymptotically hyperbolic Euclidean version (so that it has well-defined conformal boundary), even for string theory on $W^{d} \times Y^{10-d}$, where $W^{d}$ is a $d$-dimensional non-compact asymptotically hyperbolic manifold (generalizing $\text{AdS}_{d}$) and $Y^{10-d}$ a compact space (generalizing $S^{10-d}$)\cite{McInnes5}. 

\section{Topological Black Holes in General Relativity}\label{sec1}

Topological black holes can have event horizon with positive, zero, or negative scalar curvature $k$. The positively curved black holes include the usual Schwarzschild black hole with $S^{d-2}$ topology, where $d$ is the spacetime dimension, and also black holes of $S^{d-2}/\Gamma$ topology, i.e. quotient of $S^{d-2}$ by the action of some discrete group $\Gamma$. Similarly, the event horizon of $k=0$ and $k=-1$ black holes have the topology of $\Bbb{R}^{d-2}/\Gamma$ and $\Bbb{H}^{d-2}/\Gamma$, respectively.

In the context of general relativity, it was shown that $k=1$ black holes have positive brane action, while for $k=-1$ case, the brane action \emph{always} become negative and stay negative\cite{McInnes4}. Thus positively curved black holes are stable (of course being stable in Seiberg-Witten sense do not preclude the possibility that it is unstable due to other effects) but negatively curved ones are inherently unstable. Of course the onset of every instability is associated with a \emph{time scale}, thus even unstable black holes could be effectively meta-stable\cite{fbh}. 

\section{Topological Black Holes in Ho\v{r}ava-Lifshitz Gravity}

Noting that AdS/CFT correspondence is likely to occur in any quantum theory of gravity, as the existence of holographic dualities is not contingent on the validity of string theory\cite{Strominger}, we expect that something similar, if not identical, to Seiberg-Witten instability is likely to be a feature in \emph{any} quantum gravity theory that admits extended objects (e.g. branes) propagating in asymptotically AdS spaces. We therefore investigate the stability of topological black holes in the context of Ho\v{r}ava-Lifshitz Gravity\cite{yenchin2}. We find that in certain range of the detailed balance parameter $\epsilon$ (general relativity is recovered with $\epsilon=1$, while Ho\v{r}ava-Lifshitz gravity with detailed balance condition corresponds to $\epsilon=0$), the black holes in Ho\v{r}ava-Lifshitz theory can have brane action that is only negative in some \emph{finite} range of radial coordinate. This is markedly different from black holes in general relativity in which once the brane action becomes negative, it \emph{always} stays negative. Brane action with this property was previously found in the context of cosmology by Maldacena and Maoz\cite{MaldacenaMaoz}. Such black holes are expected to be stable in the sense that backreaction is very likely to set in and the systems eventually settle into a new, stable configurations. 

\section{Flat Black Holes in Einstein-Maxwell-Dilaton Theory}

The Maldacena-Maoz type of instability that occurs in Ho\v{r}ava-Lifshitz Gravity naturally raises the suspicion that it could be due to the non-relativistic and Lorentz-violating nature of Ho\v{r}ava-Lifshitz Gravity. However, we find that such instability also arises in the relatively simpler Einstein-Maxwell-Dilaton theory, which is a low energy limit of string theory. 

In particular, extending previous work in Ref.~\citen{yenchin}, we showed that\cite{yenchin3} for dilaton coupling $\alpha > 1$, asymptotically locally AdS charged dilaton Gao-Zhang black holes with flat horizon have positive brane action and thus is stable in Seiberg-Witten sense. For $0 < \alpha < 1$, the stability is of Maldacena-Maoz type. We proved that in both cases, the asymptotic behavior of the brane action is logarithmically divergent for finite $\alpha$. 

\section{Conclusion}

With the examples from Ho\v{r}ava-Lifshitz and Einstein-Maxwell-Dilaton theories, we emphasize that we still lack a \emph{quantitative} way to investigate the sufficient and necessary condition for a spacetime with Maldacena-Maoz instability capable of settling down to a stable solution due to backreaction from the brane. We know that there are manifolds where once brane action becomes negative at the conformal boundary, \emph{cannot} be deformed to have non-negative brane action\cite{McInnes5}. However, the situation in general, especially in the case of Maldacena-Maoz type of instability, is far from settled.


\begin{thebibliography}{9}

\bibitem{Mann} R. B. Mann, {\it Topological Black Holes: Outside Looking In}, Haifa 1997, Internal Structure of Black Holes and Spacetime Singularities, 311 (1997).

\bibitem{Birmingham} D. Birmingham, {\it Topological Black Holes in Anti-de Sitter Space}, Class. Quant. Grav. {\bf 16}, 1197 (1999).

\bibitem{Maldacena} J. M. Maldacena, {\it The Large N Limit of Superconformal Field Theory and Supergravity}, Adv. Theor. Math. Phys. {\bf 2}, 231 (1998). 

\bibitem{SW} N. Seiberg and E. Witten, {\it The D1/D5 System and Singular CFT}, JHEP {\bf 9904}, 017 (1999). 

\bibitem{Kleban} M. Kleban, M. Porrati, R. Rabadan, {\it Stability in Asymptotically AdS Spaces}, JHEP {\bf 0508}, 016 (2005). 

\bibitem{Barbon} Jos\'e L.F. Barb\'on, Javier Mart\'inez-Mag\'an, {\it Spontaneous Fragmentation of Topological Black Holes}, JHEP {\bf 08}, 031 (2010). 

\bibitem{MMS} J. Maldacena, J. Michelson, A. Strominger, {\it Anti-de Sitter Fragmentation}, JHEP {\bf 011}, 9902 (1999).

\bibitem{McInnes5} B. McInnes, {\it Topologically Induced Instability in String Theory}, JHEP {\bf 031}, 0103 (2001). 

\bibitem{McInnes4} B. McInnes, {\it Black Hole Final State Conspiracies}, Nucl. Phys. B {\bf 807}, 33 (2009).

\bibitem{fbh} B. McInnes, {\it Fragile Black Holes}, Nucl. Phys. B {\bf 842}, 86 (2011). 

\bibitem{Strominger} I. Bredberg, C. Keeler, V. Lysov, A. Strominger, {\it Cargese Lectures on the Kerr/CFT Correspondence}, Nucl. Phys. B, Proc. Suppl. {\bf 216}, 194 (2011).

\bibitem{yenchin2} Y. C. Ong, P. Chen, {\it Stability of Hor\v{a}va-Lifshitz Black Holes in the Context of AdS/CFT}, Phys. Rev. D {\bf 84}, 104044 (2011). 

\bibitem{MaldacenaMaoz} J. Maldacena, L. Maoz, {\it Wormholes in AdS}, JHEP {\bf 0402}, 053 (2004). 

\bibitem{yenchin} Y. C. Ong, {\it Stringy Stability of Dilaton Black Holes in 5-Dimensional Anti-de-Sitter Space}, Proceedings of the Conference in Honor of Murray Gell-Mann's 80th Birthday, 583-590, World Scientific, Singapore (2010).

\bibitem{yenchin3} Y. C. Ong, P. Chen, {\it Stringy Stability of Charged Dilaton Black Holes with Flat Event Horizon}, JHEP {\bf 08}, 079 (2012).

\end{thebibliography}
\end{document}